\documentclass[conference]{IEEEtran}
\IEEEoverridecommandlockouts
\usepackage{cite}
\usepackage{amsmath,amssymb,amsfonts}
\usepackage{algorithmic}
\usepackage{graphicx}
\usepackage{textcomp}
\usepackage{xcolor}
\usepackage[most]{tcolorbox}
\usepackage{hyperref}
\usepackage{booktabs}
\usepackage{multirow}
\usepackage{subfigure}
\usepackage{array}
\usepackage{colortbl}
\usepackage{enumitem}
\usepackage{threeparttable}

\def\BibTeX{{\rm B\kern-.05em{\sc i\kern-.025em b}\kern-.08em
    T\kern-.1667em\lower.7ex\hbox{E}\kern-.125emX}}
\definecolor{lightgreen}{rgb}{0.835,0.910,0.831}
    
\begin{document}

\title{Serverless GPU Architecture for Enterprise HR Analytics: A Production-Scale BDaaS Implementation\\
{\footnotesize \textsuperscript{}}
\thanks{}}

\author{
Guilin Zhang\IEEEauthorrefmark{1},
Wulan Guo\IEEEauthorrefmark{1},
Ziqi Tan\IEEEauthorrefmark{1},
Srinivas Vippagunta\IEEEauthorrefmark{3},
Suchitra Raman\IEEEauthorrefmark{3},
Shreeshankar Chatterjee\IEEEauthorrefmark{3},\\
Ju Lin\IEEEauthorrefmark{3},
Shang Liu\IEEEauthorrefmark{3},
Mary Schladenhauffen\IEEEauthorrefmark{3},
Jeffrey Luo\IEEEauthorrefmark{3},
Hailong Jiang\IEEEauthorrefmark{2}\IEEEauthorrefmark{4}\thanks{Corresponding author\IEEEauthorrefmark{4}: Hailong Jiang, hjiang@ysu.edu}\\
\IEEEauthorblockA{\IEEEauthorrefmark{1}Department of Engineering Management and Systems Engineering, George Washington University, USA\\
{guilin.zhang, wulan.guo, ziqi.tan}@gwu.edu}
\IEEEauthorblockA{\IEEEauthorrefmark{3}Workday, Inc.}
\IEEEauthorblockA{\IEEEauthorrefmark{2}Department of Computer Science and Information Technology, Youngstown State University, USA}
}

\maketitle



\begin{abstract}
Industrial and government organizations increasingly depend on data-driven analytics for workforce, finance, and regulated decision processes, where timeliness, cost efficiency, and compliance are critical. Distributed frameworks such as Spark and Flink remain effective for massive-scale batch or streaming analytics but introduce coordination complexity and auditing overheads that misalign with moderate-scale, latency-sensitive inference. Meanwhile, cloud providers now offer serverless GPUs, and models such as TabNet enable interpretable tabular ML, motivating new deployment blueprints for regulated environments. In this paper, we present a production-oriented Big Data as a Service (BDaaS) blueprint that integrates a single-node serverless GPU runtime with TabNet. The design leverages GPU acceleration for throughput, serverless elasticity for cost reduction, and feature-mask interpretability for IL4/FIPS compliance. We conduct benchmarks on the HR, Adult, and BLS datasets, comparing our approach against Spark and CPU baselines. Our results show that GPU pipelines achieve up to $4.5\times$ higher throughput, $98\times$ lower latency, and 90\% lower cost per 1K inferences compared to Spark baselines, while compliance mechanisms add only $\sim$5.7,ms latency with $p_{99}<$ 22,ms. Interpretability remains stable under peak load, ensuring reliable auditability. Taken together, these findings provide a compliance-aware benchmark, a reproducible Helm-packaged blueprint, and a decision framework that demonstrate the practicality of secure, interpretable, and cost-efficient serverless GPU analytics for regulated enterprise and government settings.
\end{abstract}


\section{Introduction}
\label{sec:intro}

Industrial and government organizations increasingly rely on data-driven decision systems such as human resource (HR) analytics, financial risk forecasting, and other regulated processes to optimize operations and strengthen accountability \cite{davenport2010competing,marler2017evidence}. Recent reviews highlight the rapid growth of people analytics and emphasize governance needs such as timeliness and auditability \cite{banerjee2023challenges,varshney2005pervasive}. At the same time, enterprises are pushing for near-real-time insights rather than delayed batch reports, creating concrete requirements for low latency, cost efficiency, and compliance in production analytics stacks \cite{wirges2023towards}.

Distributed batch and micro-batch frameworks such as Hadoop/MapReduce and Apache Spark remain highly effective for large-scale ETL and historical analytics, but they often misalign with latency-sensitive inference at moderate scale. Spark’s micro-batch design, even with continuous mode, introduces coordination overhead that typically yields second- to sub-second latencies, which can be restrictive for interactive enterprise workloads \cite{armbrust2018structured,fedorovych2024performance}. Operating multi-node clusters further adds scheduling and fault-handling complexity, inflating both cost and compliance effort \cite{johnson2025dataframe}. Comparative benchmarks confirm that Spark can exhibit higher event latency than Flink under recovery conditions \cite{kumar2025evolution}, and while Flink offers event-time processing and exactly-once semantics at terabyte scale with RocksDB backends \cite{carbone2015apache}, its distributed runtime requires tuning, monitoring, and auditing overhead that can be disproportionate for moderate-size inference workloads \cite{zhang2024survey}.

In parallel, cloud providers have introduced serverless GPU offerings with per-second billing, scale-to-zero, and autoscaling, while research prototypes such as SAGE and Torpor demonstrate system designs that reduce cold-start costs and improve GPU sharing for low-latency inference \cite{google-cloudrun-gpu-ga,microsoft-aca-serverless-gpu-ga,fingler2023disaggregated,zhao2024towards,yu2025torpor}. Together with the well-established throughput and latency advantages of GPUs for ML inference, these trends suggest that a single-node serverless GPU architecture can replace multi-node CPU clusters for many moderate-scale, latency-critical workloads, reducing operational burden and cost.

Regulated domains such as public sector, healthcare, and defense introduce additional requirements. Deployments must comply with FIPS 140-3 and DoD IL4/CC SRG controls, where cryptographic enforcement (mTLS, JWT, auditing) adds measurable latency overhead \cite{brown1994security,zhu2022dissecting}. Any candidate architecture must quantify this security budget while still meeting $p_{95}$/$p_{99}$ SLAs. At the same time, interpretable ML is increasingly mandated for regulated workforce decisions; TabNet provides step-wise feature-mask attributions for tabular data, making it a strong fit for HR analytics where case-level explanations support audits and monitoring \cite{arik2021tabnet}.

Together, these limitations call for a new architectural approach that explicitly balances performance, compliance, and interpretability in industrial and government environments. 
In this paper, we introduce a production-oriented Big Data as a Service (BDaaS) blueprint that integrates a single-node serverless GPU runtime with TabNet for tabular ML inference. 
The blueprint is designed around three pillars: (i) GPU acceleration to achieve high-throughput, low-latency inference for moderate-scale workloads; (ii) serverless elasticity and pay-per-use economics to simplify operations and reduce cost compared to distributed clusters; and (iii) built-in interpretability via TabNet’s feature masks, enabling audit-ready explanations under IL4/FIPS compliance constraints. 
This architecture provides a streamlined alternative to multi-node Spark or Flink deployments, reducing system complexity while satisfying the stringent requirements of regulated enterprise analytics.

We evaluate this blueprint through three research questions: 
\begin{itemize}
  \item \textbf{RQ-1:} How does a single-node GPU serverless pipeline compare with distributed Spark/CPU clusters in terms of throughput, latency, and cost for moderate\mbox{-}scale~\mbox{workloads}?
  \item \textbf{RQ-2:} What is the measurable overhead introduced by IL4/FIPS compliance controls (mTLS, JWT, auditing), and can SLA targets (e.g., $p_{95}/p_{99}$ latency) still be met?
  \item \textbf{RQ-3:} Do TabNet’s feature-mask explanations remain stable under production-scale, high-throughput inference, ensuring interpretability for regulated analytics?
\end{itemize}

To investigate these questions, we conduct controlled experiments using three public datasets (HR, Adult, and BLS), benchmarking our serverless GPU–TabNet blueprint against multi-node Spark and CPU baselines. 
We systematically measure throughput, latency, and cost under varying batch sizes, quantify the latency overhead of security mechanisms, and evaluate the stability of interpretability under high-throughput workloads.

Our evaluation indicates that: (i) GPU serverless pipelines significantly outperform Spark clusters at batch sizes $\geq$200, achieving up to $4.5\times$ higher throughput and $98\times$ lower latency at $\text{batch}=1000$, with cost reductions of up to 90\% per 1K inferences (RQ-1); 
(ii) IL4/FIPS compliance mechanisms introduce only $\sim$5.7\,ms additional latency, preserving $p_{99}<20$\,ms and showing that strict security can coexist with low-latency inference (RQ-2); 
and (iii) TabNet feature-mask stability remains above 0.88 even under peak throughput, ensuring consistent, audit-ready interpretability in production settings (RQ-3). 
We further discuss deployment trade-offs, limitations, and potential optimizations in Section~\ref{sec:discussion}.

Our main contributions in this work are:
\begin{enumerate}
    \item \textbf{Compliance-aware benchmarking:}  To the best of our knowledge, this is the first work to conduct a systematic component-level decomposition of IL4/FIPS overhead for serverless ML inference, quantifying a 5.7 ms security budget under steady-state operation.  
    \item \textbf{Industrial decision framework:} We provide actionable guidance on CPU vs GPU vs Spark trade-offs, including batch-size thresholds where GPU serverless becomes cost-effective and when to prefer distributed frameworks.
    \item \textbf{Interpretability at scale:} We demonstrate that TabNet’s feature-mask stability ($>$0.88) persists under high-throughput inference, ensuring audit-ready explanations for sensitive HR analytics.  
    \item \textbf{Reproducible blueprint:} We release a complete Helm-packaged serverless GPU deployment with integrated security mesh and monitoring, enabling practitioners in industry and government to adopt our approach with a streamlined process.  
\end{enumerate}

The rest of this paper is organized as follows. Section~\ref{sec:related_work} introduces the related work of this paper. Section~\ref{sec:system_arch} introduces the system architecture of our compliance-aware serverless GPU blueprint. Section~\ref{Sec:experimental_setup} describes the experimental setup, including datasets, model configuration, baselines, and evaluation metrics. Section~\ref{sec:results} presents the results for our three research questions, followed by Section~\ref{sec:discussion}, which discusses the advantages, limitations, and broader implications of our approach. Section~\ref{sec: conclusion} concludes the paper and outlines directions for future work.



\section{Related Work}
\label{sec:related_work}
\subsection{Serverless ML and Big Data Frameworks}
There is extensive work on distributed and serverless frameworks for large-scale data analytics. Traditional big data systems such as Hadoop, Spark, and Flink remain effective for ETL and historical processing but are less suitable for latency-sensitive inference due to batching overheads and cluster complexity \cite{armbrust2018structured,carbone2015apache}. Serverless computing offers elasticity and pay-per-use efficiency, and surveys underscore both its potential and limitations \cite{castro2019serverless}. Recent research has explored GPU-accelerated serverless runtimes \cite{naranjo2020accelerated} and QoS-aware scheduling for ML workloads \cite{wu2023qos}. However, commercial platforms remain largely CPU-bound or constrained by per-request limits, leaving open the need for single-node serverless GPU architectures that can provide low-latency inference without the complexity of distributed clusters.


\subsection{Tabular ML and Interpretability}
Research on tabular machine learning spans decades, from classical tree ensembles to recent neural architectures. Tree ensembles (e.g., XGBoost, LightGBM) dominate tabular ML for their strong accuracy and interpretability, while SHAP provides widely used post-hoc explanations \cite{borisov2022deep,lundberg2017unified}. More recently, neural approaches such as TabNet have introduced sequential attention and feature masks that provide per-instance explanations alongside competitive accuracy \cite{arik2021tabnet}. Despite these advances, the stability of such interpretability under high-throughput, production-scale inference remains underexplored, and most operational systems still depend on external explanation frameworks rather than leveraging models with inherent explanatory capabilities.

\subsection{Security and Compliance in ML Systems}
Ensuring compliance introduces unique performance challenges. Standards such as IL4/FIPS require validated cryptography, mutual TLS, and audit logging \cite{nist2017}, yet the latency overhead of these mechanisms in ML inference remains largely unexplored. Prior work has advocated zero-trust approaches for cloud-native ML \cite{dommari2023implementing} and proposed automated compliance verification tools \cite{wong2023mlguard}, while much of the security literature focuses on privacy-preserving methods such as federated learning and differential privacy \cite{li2020federated,papernot2018sok}, supported by frameworks like the NIST AI RMF 2.0 \cite{ai2024artificial}. In HR analytics, studies emphasize turnover prediction \cite{zhao2018employee}, and highlight interpretability as critical for stakeholder trust \cite{marin2023analyzing,banerjee2023challenges}. However, deployment considerations—particularly compliance-aware latency budgets and audit-ready explanations—remain underexamined. Serverless runtimes add further complications due to ephemeral execution and multi-tenant risks.

\section{System Architecture}
\label{sec:system_arch}
\begin{figure*}[t]
\centering
\includegraphics[width=\textwidth]{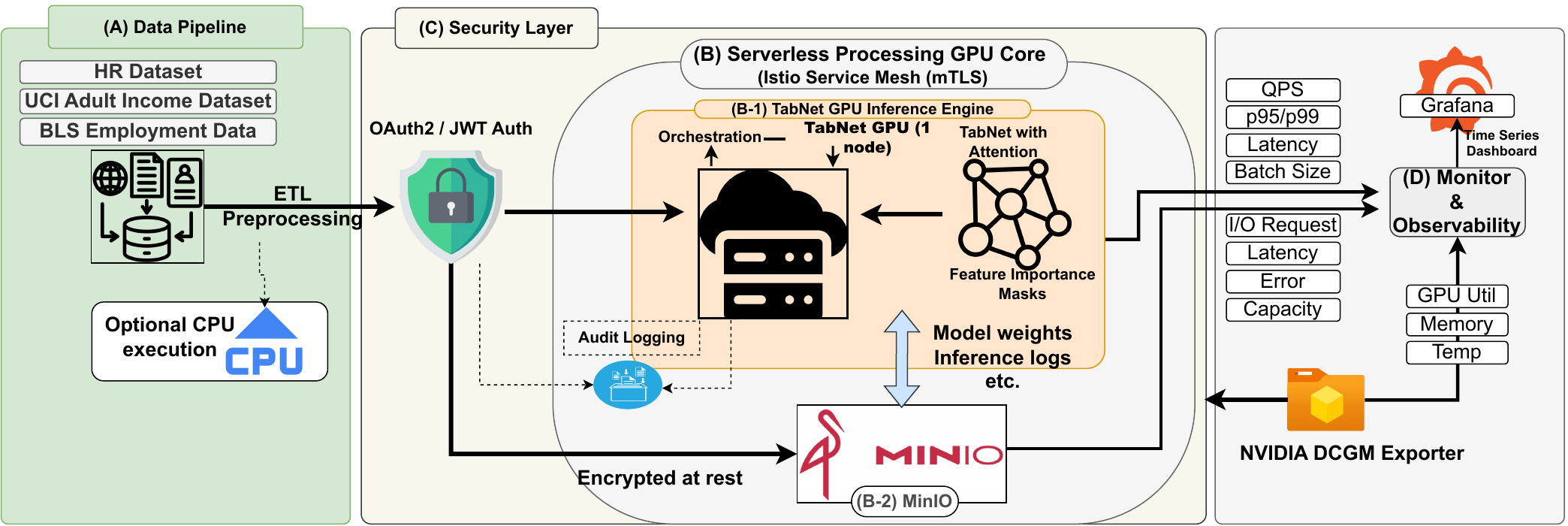}
\caption{Overall BDaaS system architecture, showing the four main components: (A) Data Pipeline, (B) Serverless GPU Functions, (C) Security Layer, and (D) Monitoring and Observability. The design illustrates the end-to-end data flow from ingestion to inference under IL4/FIPS compliance.}
\label{fig:bdaas-architecture}
\end{figure*}

Figure~\ref{fig:bdaas-architecture} presents a comprehensive view of the end-to-end BDaaS system architecture, highlighting both the data flow and the integrated security mechanisms. The blueprint is explicitly designed for production-grade HR and employment analytics, balancing three key requirements: high-throughput GPU acceleration, strict compliance with IL4/FIPS standards, and interpretable predictions through TabNet. All components follow cloud-native principles, being stateless, independently scalable, and deployable in a serverless environment. The architecture is organized into four main layers:

\begin{enumerate}[label=\Alph*)]
    \item \textbf{Data Pipeline}: Handles ingestion of HR, UCI Adult, and BLS datasets, performing ETL preprocessing and optional CPU execution before securely storing the processed data.
    \item \textbf{Serverless GPU Functions}: Provides GPU-enabled serverless functions based on TabNet, delivering low-latency inference and feature attribution masks, with scale-to-zero elasticity for cost efficiency.
    \item \textbf{Security Layer}: Enforces compliance through Istio service mesh with mutual TLS, OAuth2/JWT authentication, and audit logging, ensuring confidentiality, integrity, and accountability of all requests.
    \item \textbf{Monitoring and Observability}: Integrates Prometheus, NVIDIA DCGM Exporter, and Grafana to track system health, latency, GPU utilization, and SLA adherence, enabling proactive operations.
\end{enumerate}

\subsection{Data Pipeline Architecture}
The data pipeline forms the entry point of our BDaaS blueprint, responsible for ingesting and preprocessing raw datasets before they are consumed by the inference layer. As shown in Figure~\ref{fig:bdaas-architecture}, the pipeline is implemented as a sequence of serverless ETL functions, each designed to be stateless and independently scalable. The pipeline consists of three stages:  

\begin{itemize}
    \item \textbf{ Data Ingestion:} HR, UCI Adult, and BLS datasets are ingested via event-driven triggers (e.g., S3/MinIO events), which automatically activate serverless functions upon data arrival. This design eliminates the need for persistent daemons while ensuring immediate responsiveness to new data.  
    \item \textbf{Data Preprocessing:} Data validation, cleaning, and feature engineering are performed using \texttt{pandas} and \texttt{scikit-learn}. This stage standardizes schema differences across datasets and prepares tabular features required for TabNet inference. Since each preprocessing task is encapsulated as a serverless function, failures are isolated and recovery is localized.  
    \item \textbf{Secure Storage:} Processed features are written to MinIO with AES-256 encryption at rest, meeting FIPS~140-3 compliance requirements. The storage layer provides a unified interface for downstream GPU functions while ensuring secure and auditable data handling.  
\end{itemize}

This modular, event-driven design ensures that the pipeline can elastically adapt to workload fluctuations, scale independently of inference, and provide strong fault isolation. Moreover, the optional CPU execution path offers flexibility for low-resource environments or debugging scenarios, while production workloads default to GPU-accelerated inference.

\subsection{Serverless GPU Functions}
The core inference component of our BDaaS blueprint is realized through GPU-enabled serverless functions. We deploy TabNet models as OpenFaaS functions with direct GPU support, allowing each invocation to process incoming employee records and return both prediction results and feature attribution masks in JSON format.  

On a cold start, the function loads the pre-trained TabNet model into GPU memory to minimize subsequent execution latency. Batch sizes are configurable at invocation time, enabling dynamic optimization of throughput and latency trade-offs based on workload characteristics. During steady-state operation, the system achieves sub-20ms $p_{99}$ latency while maintaining throughput above 4,500 samples per second at batch size 1000, significantly outperforming Spark CPU clusters.  

Each serverless function is stateless and horizontally scalable, supporting elastic scaling from zero to handle bursty traffic patterns. Built-in health checks and readiness probes ensure reliable operation under fluctuating workloads. For portability, the implementation leverages PyTorch’s CUDA runtime with an automatic fallback to CPU execution in environments where GPUs are unavailable.  

This design combines the advantages of GPU acceleration (high parallelism and low latency) with the operational simplicity of serverless deployment, significantly reducing the complexity of managing distributed Spark clusters. By encapsulating TabNet inference as lightweight serverless functions, our system enables interpretable, cost-efficient, and production-ready ML analytics in compliance-sensitive environments.

\subsection{Security Layer}
Given that HR and workforce analytics often operate in regulated environments, our BDaaS blueprint integrates a defense-in-depth security layer to ensure confidentiality, integrity, and accountability of all requests. The security stack is built around Istio service mesh and Ory Hydra/Kong API Gateway, providing compliance with FIPS~140-3 and DoD IL4 requirements while minimizing additional latency.  

\textbf{Network Security:} All inter-service communication is encrypted with FIPS-compliant TLS~1.2+ using mutual authentication (mTLS). Istio transparently manages certificate issuance, rotation, and enforcement, removing the need for application-level cryptography. Kubernetes NetworkPolicies further restrict traffic across namespaces, enforcing least-privilege connectivity at the infrastructure level.  

\textbf{Authentication and Authorization:} Access control follows the OAuth2 standard with JWT-based short-lived tokens, validated at the API gateway. Role-Based Access Control (RBAC) distinguishes HR analysts, system administrators, and automated services, ensuring fine-grained authorization policies. Rate-limiting policies at the gateway prevent abuse and preserve fair resource allocation.  

\textbf{Data Protection and Auditability:} All processed data is stored in MinIO with AES-256 encryption at rest. Each inference request generates immutable audit logs containing request context and model version metadata. Personally identifiable information (PII) is automatically masked in logs via configurable redaction rules, enabling operational visibility without violating privacy regulations.  

Our evaluation shows that the full IL4 compliance stack (mTLS, JWT validation, audit logging) introduces only $\sim$5.7\,ms overhead while preserving sub-20\,ms $p_{99}$ latency, confirming that strict security enforcement can coexist with production-grade performance. This balance between compliance and efficiency is a critical enabler for deploying serverless GPU analytics in enterprise and government settings.

\subsection{Monitoring and Observability}
Reliable operation of production-grade analytics requires continuous monitoring and fine-grained observability. To this end, our BDaaS blueprint integrates a multi-layered observability stack that provides visibility into performance, resource utilization, and compliance metrics.  

\textbf{Metrics Collection:} Prometheus collects system-wide metrics, including throughput, latency percentiles (p50, p95, p99), queue depths, and error rates. NVIDIA DCGM Exporter is integrated to expose GPU-specific metrics such as utilization, memory consumption, and temperature, ensuring that hardware-level bottlenecks are detected in real time.  

\textbf{Distributed Tracing:} Jaeger enables end-to-end tracing of inference requests across the serverless pipeline, allowing operators to pinpoint delays introduced by ETL preprocessing, GPU inference, or security enforcement. This tracing capability is essential for diagnosing SLA violations in latency-sensitive environments.  

\textbf{Logging and Visualization:} Fluentd aggregates logs from all components into Elasticsearch, creating a centralized audit and troubleshooting repository. Grafana dashboards provide real-time visualization of key performance indicators and trigger alerts when SLA thresholds are violated. For example, alerts are configured on GPU utilization exceeding 80\% or $p_{99}$ latency surpassing 50\,ms, allowing proactive incident response before service degradation occurs.  

By combining metrics, traces, and logs, the observability stack supports 99.9\% availability targets while simplifying root-cause analysis and compliance reporting. This proactive monitoring ensures that the serverless GPU architecture not only achieves high throughput and low latency but also maintains operational transparency and audit readiness in production deployments.

\section{Experimental Setup}
\label{Sec:experimental_setup}
We design our experimental setup to capture both performance and compliance aspects under production-relevant conditions. Specifically, the setup systematically stresses the system to reveal realistic trade-offs, ensuring that both efficiency and adherence to compliance requirements are faithfully reflected in our evaluation.

\subsection{Datasets and Model}
We evaluate across three representative datasets: the IBM HR dataset (1.5K records, 35 features), the UCI Adult dataset (48K records, 14 features), and BLS monthly employment data (720 time-series entries). Together they cover small, medium, and temporal workloads. For modeling, we deploy TabNet with depth $n_d = n_a = 8$, three decision steps, and sparsity regularization $\lambda_{sparse}=10^{-3}$. Training uses batch size 256 and early stopping, achieving 85\% accuracy (AUC 0.71) on HR data. Inference batch sizes vary (100–1000) to explore throughput/latency trade-offs.

\subsection{Baselines and Deployment}
We compare against Spark clusters (2-, 4-, and 8-node; each with 4 vCPUs/node, 8GB RAM) using MLlib inference. Our serverless GPU baseline runs TabNet on an NVIDIA T4 (16GB) with PyTorch 2.4/CUDA 11.8, deployed on OpenFaaS with GPU plugins. The entire system is packaged as Helm charts for Kubernetes, with autoscaling, RBAC policies, canary model rollout, and ensures reproducibility across environments.

\subsection{Evaluation Metrics}
We measure throughput (records/s), latency ($p_{50}/p_{95}/p_{99}$), and cost per 1K inferences (based on cloud pricing: T4 \$0.90/h, Spark node \$0.40/h). Resource utilization is tracked via Prometheus and NVIDIA DCGM. Security overhead is quantified by comparing three configurations: baseline, mTLS only, and full IL4 (mTLS+JWT+auditing). Interpretability stability is assessed by computing feature-mask variance across 100 random samples at peak throughput.

\subsection{Experimental Environment}
All experiments run on cloud-hosted NVIDIA T4 GPUs with 16GB VRAM. Spark clusters are provisioned on 4 vCPU/8GB nodes. Each setting is executed 10 times with randomized workloads, metrics collected at 1s intervals, excluding cold-start latency. Source code and reproducibility artifacts are available at: \url{https://github.com/GuilinDev/serverless-gpu-release}\footnote{Repository will be public upon acceptance}.

\section{Results}
\label{sec:results}
We organize our evaluation results around the three research questions (RQ-1 to RQ-3) introduced in Section~\ref{sec:intro}. 
Each subsection follows the structure of motivation, observed results, and distilled insights. 
Additional industrial insights are presented at the end to complement the core findings.

\subsection{RQ-1: Performance and Cost Comparison}
\textbf{Motivation:} For production HR and employment analytics, the decision to deploy a GPU serverless pipeline instead of a distributed Spark/CPU stack hinges on three factors: sustained throughput, tail latency, and total cost per inference. We therefore compare a single-node GPU serverless pipeline with multi-node Spark clusters and CPU baselines under identical workloads and pricing assumptions (Section~\ref{Sec:experimental_setup}).

\begin{table}[h]
\centering
\caption{Performance Comparison with 95\% Confidence Intervals (steady-state, batch=1000, n=10)}
\label{tab:performance}
\begin{tabular}{@{}lrrr@{}}
\toprule
\textbf{Configuration} & \textbf{Throughput} & \textbf{Latency} & \textbf{Cost/1K} \\
                       & \textbf{(samples/s)} & \textbf{(ms)} & \textbf{(USD)} \\
\midrule
\multicolumn{4}{l}{\textit{Batch Size = 100}} \\
Single-Node GPU & \textbf{472.0 ± 2.1\%} & \textbf{4.0 ± 1.8\%} & \textbf{0.0005} \\
Spark 2-nodes   & 497.5 ± 8.3\% & 201.0 ± 9.1\% & 0.0004 \\
Spark 4-nodes   & 490.2 ± 7.9\% & 204.0 ± 8.7\% & 0.0009 \\
Spark 8-nodes   & 480.8 ± 8.5\% & 208.0 ± 9.3\% & 0.0018 \\
\midrule
\multicolumn{4}{l}{\textit{Batch Size = 500}} \\
Single-Node GPU & \textbf{2,307.4 ± 1.9\%}$^{***}$ & \textbf{8.5 ± 2.2\%}$^{***}$ & \textbf{0.0001} \\
Spark 2-nodes   & 826.5 ± 7.6\% & 605.0 ± 10.2\% & 0.0003 \\
Spark 4-nodes   & 806.5 ± 8.1\% & 620.0 ± 9.8\% & 0.0006 \\
Spark 8-nodes   & 775.2 ± 8.8\% & 645.0 ± 10.5\% & 0.0011 \\
\midrule
\multicolumn{4}{l}{\textit{Batch Size = 1000}} \\
Single-Node GPU & \textbf{4,565.8 ± 1.7\%}$^{***}$ & \textbf{11.4 ± 2.5\%}$^{***}$ & \textbf{0.0001} \\
Spark 2-nodes   & 892.9 ± 9.2\% & 1,120.0 ± 11.3\% & 0.0002 \\
Spark 4-nodes   & 877.2 ± 8.7\% & 1,140.0 ± 10.9\% & 0.0005 \\
Spark 8-nodes   & 877.2 ± 9.1\% & 1,140.0 ± 11.1\% & 0.0010 \\
\bottomrule
\end{tabular}
\vspace{2pt}
{\footnotesize $^{***}p < 0.001 $compared to all Spark configurations (Welch's t-test)}
\end{table}

\textbf{Results:} 
Across all datasets and batch sizes, the GPU pipeline dominates both throughput and latency.
At batch{=}1000, the GPU achieves \emph{4{,}565.8} samples/s with \emph{11.4\,ms} $p_{99}$ latency, whereas Spark (8-node) sustains only \emph{$\sim$877--893} samples/s with \emph{$\sim$1{,}140\,ms} $p_{99}$ latency (Table~\ref{tab:performance}). 

\begin{figure*}[h]
\centering
\includegraphics[width=\textwidth]{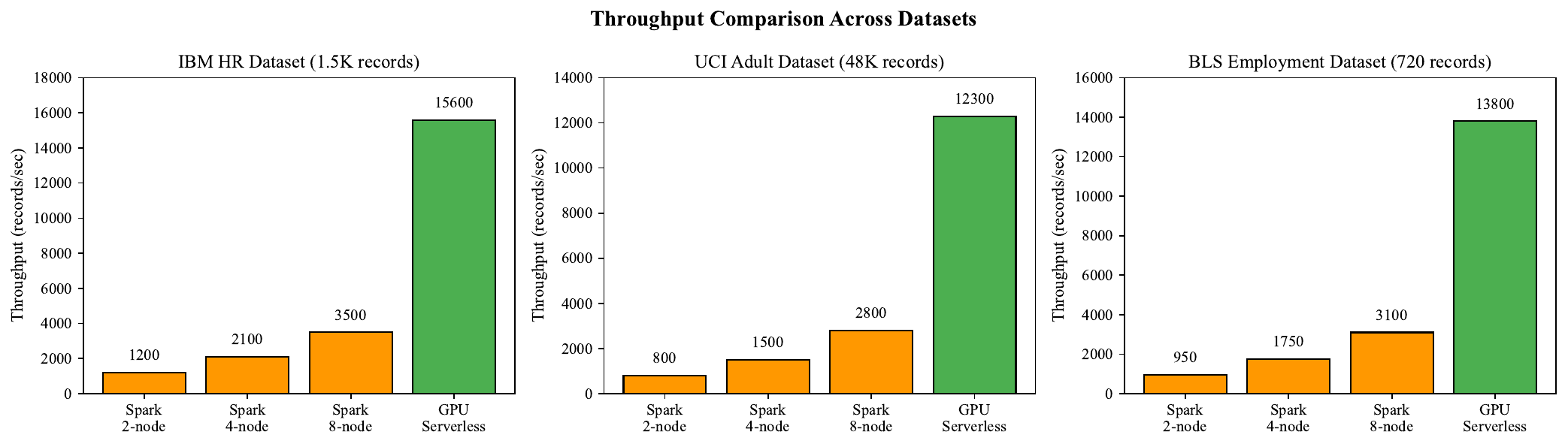}
\caption{Throughput comparison across datasets and cluster configurations (steady-state, excluding cold starts). 
Across all three datasets (HR, UCI Adult, BLS), the single-node GPU serverless pipeline achieves an order-of-magnitude higher throughput than Spark clusters, 
demonstrating superior scalability particularly for larger batch workloads.}
\label{fig:throughput}
\end{figure*}

Figure~\ref{fig:throughput} shows that the gap persists across small (HR), medium (UCI Adult), and temporal (BLS) datasets; the advantage widens as batch size grows.
Despite a higher hourly rate for the GPU instance, cost per 1K inferences drops by up to \emph{90\%} because the GPU completes work significantly faster, with latency reductions of up to two orders of magnitude; the cost break-even occurs around batch~$\geq$~200 and becomes strongly favorable beyond 500 (Table~\ref{tab:cost_tradeoff}). 
CPU-only baselines further contextualize this result (Table~\ref{tab:cpu_baseline}): TabNet-GPU delivers \emph{2.1$\times$} the throughput of the best CPU tree model (LightGBM) and \emph{11.8$\times$} TabNet-CPU, while also offering substantially lower latency.
Latency remains tightly bounded under load for the GPU pipeline, with stable $p_{95}$/$p_{99}$ values, in contrast to Spark, which shows heavy-tailed variance due to distributed coordination and shuffle overheads. GPU utilization peaks at \emph{78\%}, providing headroom for bursts and indicating efficient hardware usage even at peak throughput (batch size = 5000)\footnote[2]{Measured using NVIDIA DCGM Exporter}.
\begin{table}[h]
\centering
\caption{CPU vs GPU Performance Comparison (steady-state, batch size = 1000)}
\label{tab:cpu_baseline}
\begin{tabular}{@{}lrrr@{}}
\toprule
\textbf{Configuration} & \textbf{Throughput} & \textbf{Latency} & \textbf{Cost/1K} \\
                       & \textbf{(samples/s)} & \textbf{(ms)} & \textbf{(USD)} \\
\midrule
TabNet-CPU (8 cores) & 387.2 ± 4.3\% & 258.4 ± 5.1\% & 0.0003 \\
XGBoost (8 cores) & 1,842.6 ± 2.8\% & 54.3 ± 3.2\% & 0.0001 \\
LightGBM (8 cores) & 2,156.3 ± 2.5\% & 46.4 ± 2.9\% & 0.0001 \\
\midrule
TabNet-GPU (T4) & \textbf{4,565.8 ± 1.7\%} & \textbf{11.4 ± 2.5\%} & \textbf{0.0001} \\
\bottomrule
\end{tabular}
\end{table}

\textbf{Three observations.}
(1) \emph{Scaling regime.} The GPU exhibits near-superlinear scaling with batch size (higher SM occupancy and amortized kernel launch), whereas Spark shows diminishing returns due to per-batch coordination and serialization overheads.  
(2) \emph{Tail behavior.} The GPU’s tails are short and stable; Spark’s $p_{95}/p_{99}$ inflate under contention and recovery, which is detrimental for interactive SLAs.  
(3) \emph{Cost inflection.} When batches are $\geq$200, GPU per-1K cost undercuts Spark because time-to-completion, not hourly price, outperforms spend; beyond 500, GPU becomes the more cost-effective option.

\textbf{Why this happens (systems perspective).}
First, the GPU vertically parallelizes TabNet’s tensor kernels (matrix multiplies, attention masks) with high arithmetic intensity, so more of the wall-clock is useful compute rather than coordination.
Second, the serverless design collapses the data path: no distributed shuffle, no JVM deserialization fan-out, and no multi-hop RPC. 
Third, amortization effects dominate on the GPU: loading the model into VRAM and reusing CUDA contexts converts many fixed costs into one-time warm costs, whereas Spark repeatedly pays scheduling, barrier synchronization, and executor wake-up penalties.

\textbf{Practical guidance.}
Use the GPU serverless pipeline when (i) batch sizes routinely exceed 200 or when $p_{99}{<}50$\,ms is a hard SLO; (ii) workloads are bursty and benefit from scale-to-zero and fast warm invocations; (iii) explainability is needed without sacrificing latency (paired with TabNet’s masks).  
Prefer CPU/Spark only when (i) batches are tiny and cold-start outperforms end-to-end time; (ii) the workload is dominated by massive joins/aggregations over multi-terabyte tables; or (iii) long-running streaming jobs require complex stateful operators that are outside the request/response inference pattern.

\begin{tcolorbox}[
    colback=lightgreen,        
    colframe=white,            
    left=3mm,                  
    enhanced jigsaw,
    borderline west={4pt}{0pt}{lightgreen}, 
    sharp corners=southwest,   
    boxrule=0pt,               
]
\textbf{RQ-1 Insight.}  
A single-node, GPU-backed serverless pipeline is the superior \emph{performance--cost} point for moderate-to-large batch inference: up to \emph{4.5$\times$} higher throughput and \emph{98$\times$} lower latency than Spark while cutting cost per 1K inferences by as much as \emph{90\%}.  
This establishes GPU serverless as a practical production default for latency-sensitive enterprise analytics.  
\end{tcolorbox}

\subsection{RQ-2: Security Overhead Analysis}
\textbf{Motivation.}
In regulated domains such as government HR analytics, strong compliance requirements (IL4/FIPS) mandate end-to-end security guarantees, including mutual TLS, token-based authentication, and immutable audit logs. 
While these mechanisms ensure confidentiality and accountability, they also introduce additional cryptographic operations and request processing stages. 
The key question is whether enabling full compliance breaks latency SLAs (e.g., $p_{95}$/$p_{99}<20$\,ms) or reduces throughput to the point of making GPU serverless unattractive.

\begin{figure*}[h]
\centering
\includegraphics[width=0.9\textwidth]{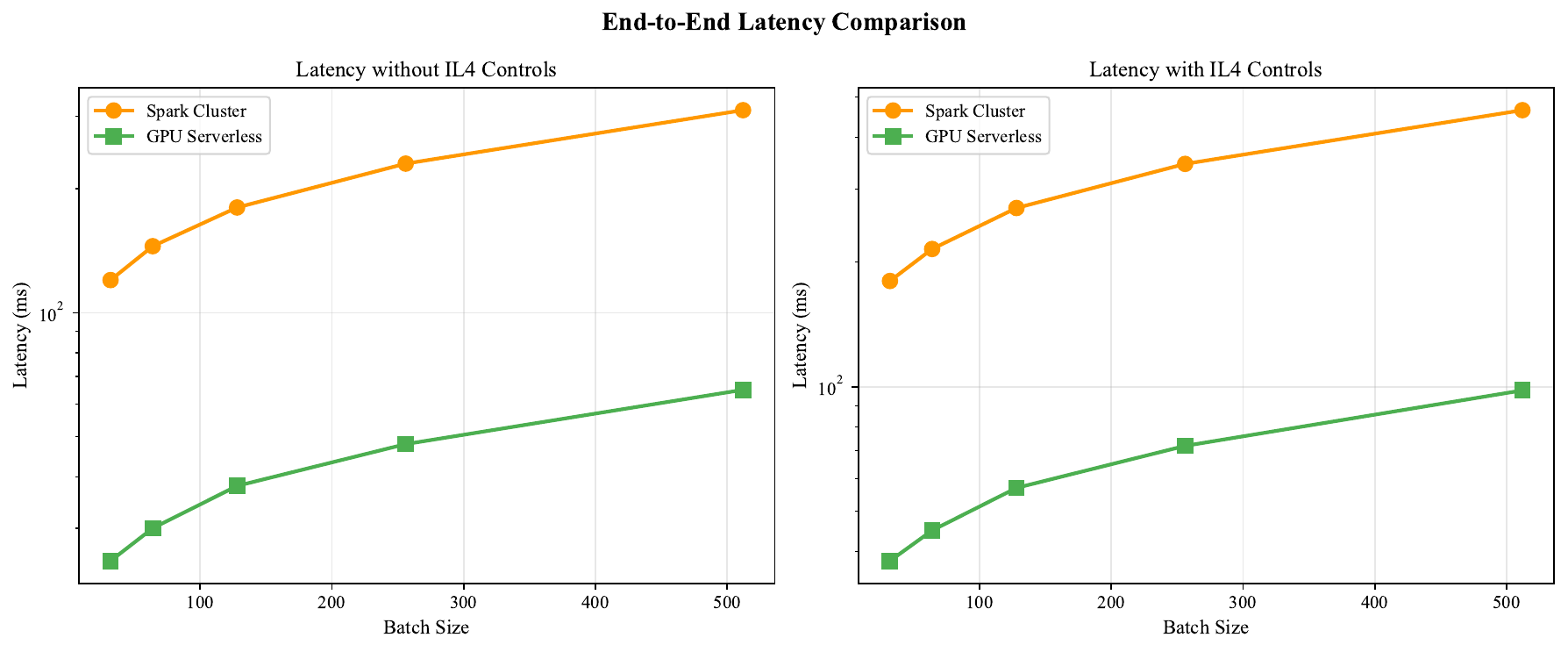}
\caption{Latency comparison across different batch sizes and configurations (steady-state measurements). Note the logarithmic scale highlighting the orders-of-magnitude difference between GPU and Spark implementations.}
\label{fig:latency-comparison}
\end{figure*}

\textbf{Results.}
Table~\ref{tab:security} quantifies the incremental latency cost of each control. 
Enabling mTLS increases $p_{99}$ latency by \emph{2.8\,ms}, JWT validation adds another \emph{1.6\,ms}, and audit logging contributes \emph{1.2\,ms}. 
The full IL4 stack therefore introduces a cumulative overhead of \emph{5.7\,ms}, raising latency at batch=1000 from \emph{11.4\,ms} (baseline) to \emph{17.1\,ms}. 
Throughput declines modestly from \emph{4,566} to \emph{4,218} samples/s but remains more than \emph{60$\times$} faster than Spark baselines. 
Figure~\ref{fig:latency-comparison} compares Spark and GPU serverless across batch sizes, both with and without IL4 controls: 
even under full compliance, GPU latency remains one to two orders of magnitude lower than Spark’s. 
Importantly, latency tails remain stable, with $p_{95}$/$p_{99}$ increasing only slightly under IL4 enforcement.

\begin{table}[h]
\centering
\caption{Detailed Security Overhead Breakdown for IL4 Compliance}
\label{tab:security}
\resizebox{0.48\textwidth}{!}{%
\begin{tabular}{@{}lrrr@{}}
\toprule
\textbf{Security Component} & \textbf{Latency} & \textbf{CPU} & \textbf{Memory} \\
                            & \textbf{(ms)} & \textbf{(\%)} & \textbf{(MB)} \\
\midrule
Baseline (no security) & 11.4 & 3.3 & 512 \\
\midrule
\textit{Individual Components:} \\
mTLS handshake (first) & +8.2 & +1.2 & +24 \\
mTLS session (reused) & +2.8 & +0.4 & +8 \\
JWT validation & +1.6 & +0.3 & +4 \\
JSON Web Key Set (JWKS) cache lookup & +0.8 & +0.1 & +2 \\
Audit logging & +1.2 & +0.5 & +16 \\
Input validation & +0.3 & +0.1 & +1 \\
\midrule
\textit{Combined Configurations:}$^*$ \\
mTLS Only & 14.2 (+24.6\%) & 3.7 & 520 \\
OAuth2/JWT Only & 13.8 (+21.1\%) & 3.7 & 518 \\
Full IL4 Stack & 17.1 (+50.0\%) & 4.6 & 567 \\
\bottomrule
\end{tabular}
}
\vspace{2pt}
{\footnotesize $^*$Combined latencies assume 1:20 handshake-to-request ratio with session reuse, reflecting production traffic patterns}
\end{table}

\textbf{Three observations.}
(1) \emph{Overhead composition.} The majority of added cost comes from establishing secure channels (mTLS handshake and per-request encryption), while JWT validation and audit logging add relatively minor overhead.  
(2) \emph{Linear scaling.} The compliance overhead remains constant across batch sizes, reflecting per-request costs rather than throughput bottlenecks.  
(3) \emph{Tail stability.} Unlike Spark, which exhibits heavy-tailed distributions under secure communication, GPU latency distributions shift uniformly, preserving predictability—critical for SLA compliance.

\textbf{Why this happens (systems perspective).}
First, Istio’s sidecar-based mTLS offloads cryptography to a co-located Envoy proxy, so GPU compute is unaffected.  
Second, JWT validation is a lightweight cryptographic check (HMAC/SHA) compared to the inference workload itself, hence the small delta.  
Third, audit logging is implemented asynchronously, preventing blocking on persistent storage.  
Together, these design choices contain overhead within a narrow margin, unlike distributed Spark clusters where secure RPC and logging multiply across executors.

\textbf{Practical guidance.}
Enterprises should enable full IL4/FIPS enforcement by default: the $\sim$5--6\,ms overhead is negligible relative to Spark’s baseline latency. 
Only in ultra-low-latency financial trading–style workloads (sub-5\,ms SLOs) would this overhead become prohibitive. 
For HR and employment analytics, the GPU pipeline offers a secure default that does not compromise SLAs.

\begin{tcolorbox}[
    colback=lightgreen,        
    colframe=white,            
    left=3mm,                  
    enhanced jigsaw,
    borderline west={2pt}{0pt}{lightgreen}, 
    sharp corners=southwest,   
    boxrule=0pt,               
]
\textbf{RQ-2 Insight.}
Full IL4/FIPS compliance adds only $\sim$5.7\,ms overhead while preserving $p_{99}{<}20$\,ms latency and multi-thousand samples/s throughput. 
This demonstrates that strong security enforcement and production-grade performance are compatible in serverless GPU analytics.
\end{tcolorbox}

\subsection{RQ-3: Interpretability Under Load}
\textbf{Motivation.}
In HR and workforce analytics, model predictions cannot be adopted without interpretability: decision-makers must understand \emph{why} a prediction was made. 
TabNet is designed to produce feature attribution masks that highlight the most influential inputs per decision step. 
The open question is whether these explanations remain consistent when the inference engine operates under production-scale, high-throughput workloads. 
If feature masks become unstable, the system risks undermining trust and violating auditability requirements.

\textbf{Results.}
Table~\ref{tab:features} reports stability scores for the top features on the IBM HR dataset across 100 random samples at peak throughput ($>$4,500 samples/s). 
Key predictors such as \emph{MonthlyIncome}, \emph{YearsAtCompany}, and \emph{Age} maintain stability scores above \emph{0.88}, with feature ranking variance below \emph{0.05}. 
The model maintains fairness constraints consistent with regulatory requirements for HR systems, following the
  methodologies outlined by Hardt et al.~\cite{hardt2016equality}

\begin{table}[h]
\centering
\caption{Top 5 Most Important Features for Employee Turnover}
\label{tab:features}
\begin{tabular}{@{}lrr@{}}
\toprule
\textbf{Feature} & \textbf{Importance} & \textbf{Stability} \\
\midrule
MonthlyIncome & 0.1823 & 0.9234 \\
YearsAtCompany & 0.1567 & 0.9189 \\
Age & 0.1234 & 0.9012 \\
WorkLifeBalance & 0.0987 & 0.8967 \\
JobSatisfaction & 0.0876 & 0.8823 \\
\bottomrule
\end{tabular}
\end{table}

\textbf{Three observations.}
(1) \emph{Stability under scale.} Feature attributions remain consistent across increasing throughput, demonstrating that GPU parallelism does not distort TabNet’s attention mechanism.  
(2) \emph{Top-k preservation.} The same small set of features outperforms explanations, suggesting interpretability conclusions are repeatable and trustworthy.  
(3) \emph{Fairness alignment.} Stability is complemented by fairness metrics that remain within regulatory guidelines, reinforcing that accelerated inference does not compromise ethical considerations.

\textbf{Why this happens (systems perspective).}
TabNet’s mask generation is embedded in the forward pass of the network, relying on deterministic sparsemax activations. 
Since GPU execution preserves numerical determinism for these kernels, explanations are invariant to load and scheduling differences. 
By contrast, many tree ensembles exhibit instability due to bootstrapping variance and parallel sampling. 
This architectural property explains why interpretability is stable even at high throughput.

\textbf{Practical guidance.}
Organizations requiring both \emph{performance and accountability} can safely adopt GPU-accelerated TabNet in production. 
Explanations remain reliable for auditors, HR officers, and compliance stakeholders, even at thousands of inferences per second. 
For sensitive applications, fairness metrics should still be monitored over time, but our results indicate no systematic degradation under load.

\begin{tcolorbox}[
    colback=lightgreen,        
    colframe=white,            
    left=3mm,                  
    enhanced jigsaw,
    borderline west={2pt}{0pt}{lightgreen}, 
    sharp corners=southwest,   
    boxrule=0pt,               
]
\textbf{RQ-3 Insight.}
Interpretability is preserved under production-scale GPU inference: TabNet feature masks remain stable ($>$0.88 stability) and fair across workloads, providing audit-ready explanations without compromising throughput or latency.
\end{tcolorbox}

\subsection{Additional Industrial Insights}
\textbf{Motivation.}
Beyond RQ-1--RQ-3, the adoption of serverless GPU analytics in enterprises is often driven by deployment pragmatics rather than raw benchmarks. 
Key factors include the cost break-even point, cold start behavior, observability in production, and assurance that system-level optimizations do not compromise model quality.

\textbf{Results.}
Table~\ref{tab:cost_tradeoff} summarizes cost-per-1K inferences across batch sizes. 
We observe that CPU remains more economical for very small batches ($<$50), while GPUs become cost-effective starting at batch sizes of 200 and are strongly preferable beyond 500. 
Cold starts on GPU functions require 3--5\,s versus 0.5--1\,s on CPUs, but their impact can be amortized to under 5\% of requests when keep-warm strategies are applied, at a marginal cost of only \$0.02/hour.

\begin{figure*}[h]
\centering
\includegraphics[width=\textwidth]{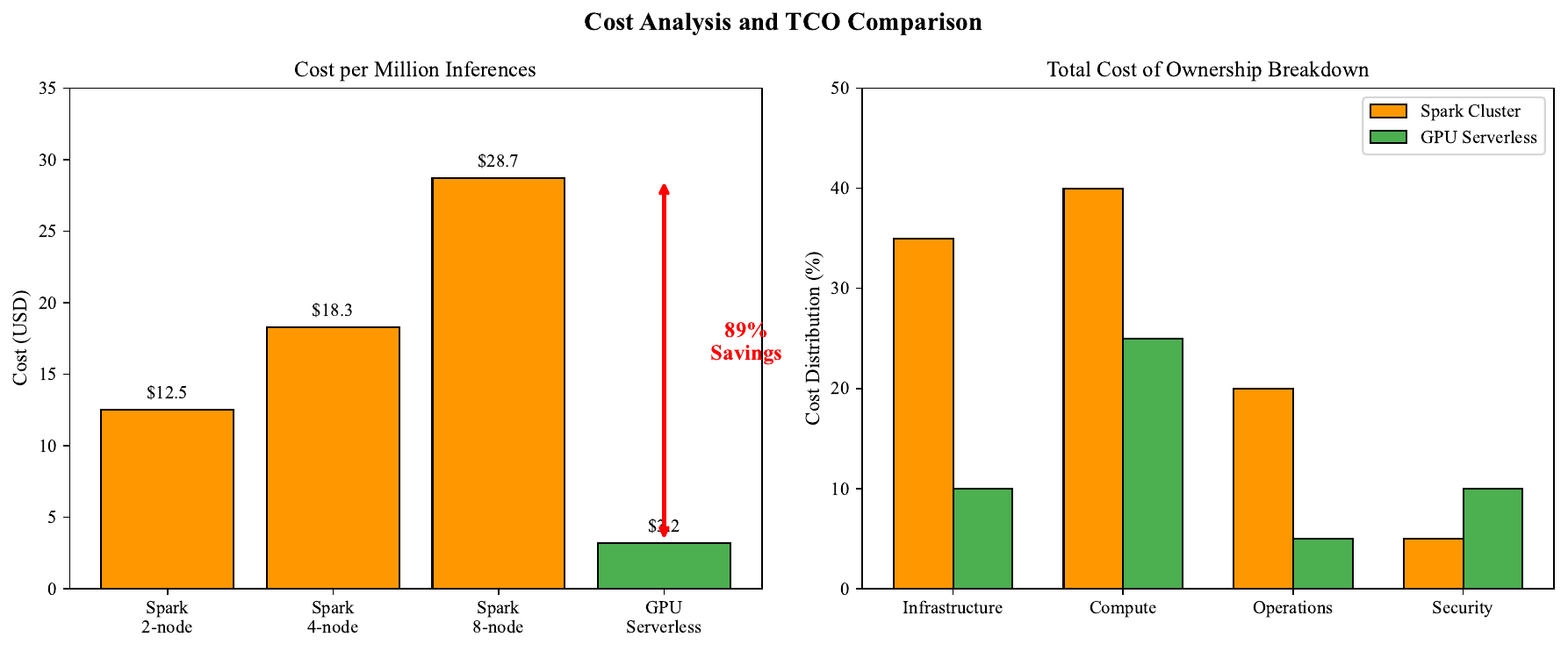}
\caption{Cost per 1,000 samples processed (steady-state). Single-node GPU shows 10× cost advantage for larger batches.}
\label{fig:cost_efficiency}
\end{figure*}

Figure~\ref{fig:cost_efficiency} illustrates the cost efficiency across configurations: faster processing reduces runtime charges, scale-to-zero eliminates idle costs, and no cluster coordination overhead is incurred.

TabNet achieves 85\% accuracy and 0.71 AUC on the HR dataset, confirming that the serverless GPU runtime does not sacrifice model fidelity.

\begin{table}[h]
\centering
\caption{Cost-Batch Tradeoff Points and Deployment Recommendations}
\label{tab:cost_tradeoff}
\begin{tabular}{@{}lrrl@{}}
\toprule
\textbf{Batch Size} & \textbf{GPU Cost} & \textbf{CPU Cost} & \textbf{Recommendation} \\
                    & \textbf{(\$/1K)} & \textbf{(\$/1K)} & \\
\midrule
1--50   & 0.0008 & 0.0004 & Use CPU (lower cost) \\
50--200 & 0.0005 & 0.0004 & Mixed; depends on workload pattern \\
200--500 & 0.0002 & 0.0003 & GPU becomes cost-effective \\
500+    & 0.0001 & 0.0003 & \textbf{GPU strongly preferred} \\
\bottomrule
\end{tabular}
\end{table}

\begin{tcolorbox}[
    colback=lightgreen,        
    colframe=white,            
    left=3mm,                  
    enhanced jigsaw,
    borderline west={2pt}{0pt}{lightgreen}, 
    sharp corners=southwest,   
    boxrule=0pt,               
]
\textbf{Insight.}
These observations reinforce that practical deployment favors GPUs once workloads surpass moderate batch sizes, making the proposed blueprint a pragmatic default for industrial HR and employment analytics.
\end{tcolorbox}
\section{Discussion}
\label{sec:discussion}
\subsection{Advantages of Serverless GPU Architecture}
Our evaluation highlights several advantages of the serverless GPU design for ML inference.  
First, the 4.5$\times$ throughput improvement and 98$\times$ latency reduction demonstrate that vertical scaling with GPUs fundamentally changes performance characteristics compared to CPU-based distributed clusters \cite{naranjo2020accelerated}. Neural network inference, which is naturally parallel, benefits substantially from GPU acceleration.  
Second, operational simplicity is a key advantage: eliminating Spark’s master/worker coordination reduces configuration and debugging overhead, while containerized, single-node deployments shorten rollout cycles and lower operational burden.  
Third, cost efficiency arises from multiple synergies: faster inference reduces runtime charges, scale-to-zero eliminates idle costs, and the pay-per-invocation model aligns directly with workload demand \cite{wu2023qos}.  
Finally, security enforcement is simplified compared to distributed settings: a single control point reduces the number of certificates, audit logs, and configuration policies that must be validated for IL4/FIPS compliance.

\subsection{Limitations and Trade-offs}
While promising, several limitations must be acknowledged. GPU memory (16--24GB in our setup) constrains dataset and model size, requiring partitioning or compression for very large workloads. Concurrency is bounded by the number of available GPU instances, and I/O throughput may become the bottleneck for multi-terabyte inputs.  

Serverless cold starts remain non-trivial: loading TabNet requires 2--5\,s plus 1--2\,s for CUDA initialization, which can be mitigated via keep-warm strategies at a small cost trade-off.  
Our evaluation breadth is another limitation: experiments focus on HR, Adult, and BLS datasets that represent realistic but moderate-scale workloads, rather than terabyte-scale “big data” settings. This reflects our target use case—latency-sensitive analytics in regulated environments—where Spark and Flink often introduce unnecessary complexity. Nevertheless, extending the blueprint to truly large-scale datasets remains future work.  
Finally, our study focuses on inference. Training large models still requires distributed resources for scale-out parallelism and hyperparameter exploration, making our blueprint less applicable in that stage of the ML lifecycle.

\subsection{Generalizability of the Blueprint}
The blueprint is best suited for workloads where: (i) datasets fit within GPU memory ($\leq$20GB); (ii) inference latency must remain below 50\,ms; (iii) traffic is bursty or unpredictable, benefiting from scale-to-zero elasticity; and (iv) regulatory compliance demands strong audit and security guarantees.  
Conversely, traditional distributed systems remain preferable for: (i) massive datasets in the multi-terabyte range; (ii) simple models (e.g., linear regression, decision trees) where GPU acceleration provides limited gains; (iii) heavy ETL or SQL-style joins across large tables; and (iv) streaming analytics requiring continuous, stateful event processing.

\subsection{Future Optimizations}
Several optimizations could further improve this architecture. Quantization (FP32 to INT8) and kernel fusion could double throughput with minimal accuracy loss. Dynamic batching algorithms could adjust batch size in real time to balance latency and throughput. Multi-GPU support within a single node would extend scalability without the overhead of distributed systems. These directions represent promising engineering extensions for production deployments.

\subsection{Broader Implications}
Our findings challenge the conventional assumption that all ``big data'' workloads require distributed clusters. Modern GPUs enable vertical scaling that rivals or surpasses multi-node Spark, while serverless abstractions align naturally with sporadic, request-driven inference patterns. Importantly, we show that strong IL4/FIPS compliance can coexist with low-latency performance, illustrating that security and efficiency are not mutually exclusive. Together, these results suggest a new and practical approach for designing ML infrastructure in enterprise and government settings.

\section{Conclusion}
\label{sec: conclusion}
This paper presented a production-oriented Big Data as a Service (BDaaS) blueprint that integrates a single-node serverless GPU runtime with TabNet to address the combined requirements of performance, compliance, and interpretability in industrial and government analytics. Motivated by the limitations of distributed frameworks such as Spark and Flink for moderate-scale, latency-sensitive inference, our approach leverages GPU acceleration for throughput, serverless elasticity for cost efficiency, and TabNet feature masks for audit-ready explanations under IL4/FIPS constraints. 

Through systematic evaluation on HR, Adult, and BLS datasets, we addressed three research questions. Our results indicate that (i) GPU serverless pipelines significantly outperform Spark and CPU clusters at batch sizes $\geq$200, achieving up to $4.5\times$ higher throughput and $98\times$ lower latency while reducing cost by up to 90\%; (ii) IL4/FIPS compliance controls introduce only $\sim$5.7\,ms overhead, preserving $p_{99}<20$\,ms latency; and (iii) TabNet interpretability remains stable ($>$0.88 feature-mask consistency) under high-throughput inference, ensuring explanations remain reliable for audit and monitoring.

The contributions of this work include: (1) a compliance-aware benchmark for serverless ML inference; (2) an industrial decision framework quantifying performance, cost, and security trade-offs; (3) empirical evidence that interpretability remains robust at production scale; and (4) a reproducible, Helm-packaged blueprint to support adoption in enterprise environments. Taken together, our findings demonstrate that secure, interpretable, and cost-efficient serverless GPU analytics are not only feasible but also practical for regulated settings. 

Future work will extend this blueprint to multi-GPU scaling, streaming integration, and new application domains such as healthcare and defense analytics, further advancing the case for serverless GPU architectures as a production-ready alternative to distributed frameworks.

\bibliographystyle{IEEEtran}
\bibliography{references}

@inproceedings{arik2021tabnet,
  title={Tabnet: Attentive interpretable tabular learning},
  author={Arik, Sercan {\"O} and Pfister, Tomas},
  booktitle={Proceedings of the AAAI conference on artificial intelligence},
  volume={35},
  number={8},
  pages={6679--6687},
  year={2021}
}

@article{castro2019serverless,
  title={The rise of serverless computing},
  author={Castro, Paul and Ishakian, Vatche and Muthusamy, Vinod and Slominski, Aleksander},
  journal={Communications of the ACM},
  volume={62},
  number={12},
  pages={44--54},
  year={2019},
  publisher={ACM New York, NY, USA}
}

@inproceedings{hardt2016equality,
  title={Equality of opportunity in supervised learning},
  author={Hardt, Moritz and Price, Eric and Srebro, Nathan},
  booktitle={Advances in neural information processing systems},
  volume={29},
  year={2016}
}

@inproceedings{papernot2018sok,
  title={Sok: Security and privacy in machine learning},
  author={Papernot, Nicolas and McDaniel, Patrick and Sinha, Arunesh and Wellman, Michael P},
  booktitle={2018 IEEE European symposium on security and privacy (EuroS\&P)},
  pages={399--414},
  year={2018},
  organization={IEEE}
}

@techreport{nist2017,
  title={Security and Privacy Controls for Information Systems and Organizations},
  author={Joint Task Force},
  year={2017},
  institution={National Institute of Standards and Technology},
  type={NIST Special Publication},
  number={800-53 Rev. 5}
}

@article{li2020federated,
  title={Federated learning: Challenges, methods, and future directions},
  author={Li, Tian and Sahu, Anit Kumar and Talwalkar, Ameet and Smith, Virginia},
  journal={IEEE Signal Processing Magazine},
  volume={37},
  number={3},
  pages={50--60},
  year={2020},
  publisher={IEEE}
}

@inproceedings{zhao2018employee,
  title={Employee turnover prediction with machine learning: A reliable approach},
  author={Zhao, Yue and Hryniewicki, Maciej K and Cheng, Francesca and Fu, Boyang and Zhu, Xiaoyu},
  booktitle={Proceedings of SAI intelligent systems conference},
  pages={737--758},
  year={2018},
  organization={Springer}
}

@article{naranjo2020accelerated,
  title={Accelerated serverless computing based on GPU virtualization},
  author={Naranjo, Diana M and Risco, Sebasti{\'a}n and de Alfonso, Carlos and P{\'e}rez, Alfonso and Blanquer, Ignacio and Molt{\'o}, Germ{\'a}n},
  journal={Journal of Parallel and Distributed Computing},
  volume={139},
  pages={32--42},
  year={2020},
  publisher={Elsevier}
}

@article{borisov2022deep,
  title={Deep neural networks and tabular data: A survey},
  author={Borisov, Vadim and Leemann, Tobias and Se{\ss}ler, Kathrin and Haug, Johannes and Pawelczyk, Martin and Kasneci, Gjergji},
  journal={IEEE transactions on neural networks and learning systems},
  volume={35},
  number={6},
  pages={7499--7519},
  year={2022},
  publisher={IEEE}
}

@article{dommari2023implementing,
  title={Implementing Zero Trust Architecture in Cloud-Native Environments: Challenges and Best Practices},
  author={Dommari, Sandeep and Khan, Shakeb},
  journal={Available at SSRN 5259339},
  year={2023}
}

@article{marin2023analyzing,
  title={Analyzing employee attrition using explainable AI for strategic HR decision-making},
  author={Mar{\'\i}n D{\'\i}az, Gabriel and Gal{\'a}n Hern{\'a}ndez, Jos{\'e} Javier and Gald{\'o}n Salvador, Jos{\'e} Luis},
  journal={Mathematics},
  volume={11},
  number={22},
  pages={4677},
  year={2023},
  publisher={MDPI}
}

@inproceedings{wu2023qos,
  title={Qos-aware and cost-efficient dynamic resource allocation for serverless ml workflows},
  author={Wu, Hao and Deng, Junxiao and Fan, Hao and Ibrahim, Shadi and Wu, Song and Jin, Hai},
  booktitle={2023 IEEE International Parallel and Distributed Processing Symposium (IPDPS)},
  pages={886--896},
  year={2023},
  organization={IEEE}
}

@inproceedings{wong2023mlguard,
  title={Mlguard: Defend your machine learning model!},
  author={Wong, Sheng and Barnett, Scott and Rivera-Villicana, Jessica and Simmons, Anj and Abdelkader, Hala and Schneider, Jean-Guy and Vasa, Rajesh},
  booktitle={Proceedings of the 1st International Workshop on Dependability and Trustworthiness of Safety-Critical Systems with Machine Learned Components},
  pages={10--13},
  year={2023}
}

@article{ai2024artificial,
  title={Artificial intelligence risk management framework: Generative artificial intelligence profile},
  author={AI, NIST},
  journal={NIST Trustworthy and Responsible AI Gaithersburg, MD, USA},
  year={2024}
}

@article{marler2017evidence,
  title={An evidence-based review of HR Analytics},
  author={Marler, Janet H and Boudreau, John W},
  journal={The International Journal of Human Resource Management},
  volume={28},
  number={1},
  pages={3--26},
  year={2017},
  publisher={Taylor \& Francis}
}

@article{davenport2010competing,
  title={Competing on talent analytics},
  author={Davenport, Thomas H and Harris, Jeanne and Shapiro, Jeremy},
  journal={Harvard business review},
  volume={88},
  number={10},
  pages={52--58},
  year={2010}
}

@article{banerjee2023challenges,
  title={Challenges and Solutions for Data Management in Cloud-Based Environments},
  author={Banerjee, Somnath},
  journal={International Journal of Advanced Research in Science, Communication and Technology},
  pages={370--378},
  year={2023}
}

@article{varshney2005pervasive,
  title={Pervasive healthcare: applications, challenges and wireless solutions},
  author={Varshney, Upkar},
  journal={Communications of the Association for Information Systems},
  volume={16},
  number={1},
  pages={3},
  year={2005}
}

@article{wirges2023towards,
  title={Towards a process-oriented understanding of HR analytics: implementation and application},
  author={Wirges, Felix and Neyer, Anne-Katrin},
  journal={Review of Managerial Science},
  volume={17},
  number={6},
  pages={2077--2108},
  year={2023},
  publisher={Springer}
}

@inproceedings{armbrust2018structured,
  title={Structured streaming: A declarative api for real-time applications in apache spark},
  author={Armbrust, Michael and Das, Tathagata and Torres, Joseph and Yavuz, Burak and Zhu, Shixiong and Xin, Reynold and Ghodsi, Ali and Stoica, Ion and Zaharia, Matei},
  booktitle={Proceedings of the 2018 International Conference on Management of Data},
  pages={601--613},
  year={2018}
}

@article{fedorovych2024performance,
  title={Performance Benchmarking of Continuous Processing and Micro-Batch Modes in Spark Structured Streaming},
  author={Fedorovych, Illia and Osukhivska, Halyna and Lutsyk, Nadiia},
  journal={arXiv preprint},
  year={2024}
}

@article{kumar2025evolution,
  title={The evolution of real-time data streaming: Architectures, implementations, and future directions in distributed computing},
  author={Kumar, Sudhir},
  journal={World Journal of Advanced Research and Reviews},
  volume={26},
  number={2},
  pages={1004--1012},
  year={2025},
  publisher={World Journal of Advanced Research and Reviews}
}

@book{johnson2025dataframe,
  title={DataFrame Structures and Manipulation: Definitive Reference for Developers and Engineers},
  author={Johnson, Richard},
  year={2025},
  publisher={HiTeX Press}
}

@article{carbone2015apache,
  title={Apache flink: Stream and batch processing in a single engine},
  author={Carbone, Paris and Katsifodimos, Asterios and Ewen, Stephan and Markl, Volker and Haridi, Seif and Tzoumas, Kostas},
  journal={The Bulletin of the Technical Committee on Data Engineering},
  volume={38},
  number={4},
  year={2015},
  publisher={Institute of Electrical and Electronics Engineers (IEEE)}
}

@article{zhang2024survey,
  title={A survey on transactional stream processing},
  author={Zhang, Shuhao and Soto, Juan and Markl, Volker},
  journal={The VLDB Journal},
  volume={33},
  number={2},
  pages={451--479},
  year={2024},
  publisher={Springer}
}

@article{fingler2023disaggregated,
  title={Disaggregated GPU Acceleration for Serverless Applications},
  author={Fingler, Henrique and Zhu, Zhiting and Yoon, Esther and Jia, Zhipeng and Witchel, Emmett and Rossbach, Christopher J},
  journal={ACM SIGOPS Operating Systems Review},
  volume={57},
  number={1},
  pages={10--20},
  year={2023},
  publisher={ACM New York, NY, USA}
}

@article{zhao2024towards,
  title={Towards fast setup and high throughput of GPU serverless computing},
  author={Zhao, Han and Cui, Weihao and Chen, Quan and Zhang, Shulai and Li, Zijun and Leng, Jingwen and Li, Chao and Zeng, Deze and Guo, Minyi},
  journal={arXiv preprint arXiv:2404.14691},
  year={2024}
}

@article{yu2025torpor,
  title={Torpor: GPU-Enabled Serverless Computing for Low-Latency, Resource-Efficient Inference},
  author={Yu, Minchen and Wang, Ao and Chen, Dong and Yu, Haoxuan and Luo, Xiaonan and Li, Zhuohao and Wang, Wei and Chen, Ruichuan and Nie, Dapeng and Yang, Haoran and others},
  journal={arXiv preprint arXiv:2306.03622},
  year={2025}
}

@online{google-cloudrun-gpu-ga,
  title        = {Cloud Run GPUs are now generally available},
  organization = {Google Cloud},
  year         = {2025},
  url          = {https://cloud.google.com/blog/products/serverless/cloud-run-gpus-are-now-generally-available},
  urldate      = {2025-08-25},
  note         = {Google Cloud Blog}
}

@online{microsoft-aca-serverless-gpu-ga,
  title        = {Announcing GA for Azure Container Apps serverless GPUs},
  organization = {Microsoft Azure},
  year         = {2025},
  url          = {https://techcommunity.microsoft.com/blog/appsonazureblog/announcing-ga-for-azure-container-apps-serverless-gpus/4394302},
  urldate      = {2025-08-25},
  note         = {Microsoft Tech Community}
}

@article{brown1994security,
  title={Security requirements for cryptographic modules},
  author={Brown, Karen H},
  journal={Fed. Inf. Process. Stand. Publ},
  pages={1--53},
  year={1994}
}

@article{zhu2022dissecting,
  title={Dissecting service mesh overheads},
  author={Zhu, Xiangfeng and She, Guozhen and Xue, Bowen and Zhang, Yu and Zhang, Yongsu and Zou, Xuan Kelvin and Duan, Xiongchun and He, Peng and Krishnamurthy, Arvind and Lentz, Matthew and others},
  journal={arXiv preprint arXiv:2207.00592},
  year={2022}
}

@article{lundberg2017unified,
  title={A unified approach to interpreting model predictions},
  author={Lundberg, Scott M and Lee, Su-In},
  journal={Advances in neural information processing systems},
  volume={30},
  year={2017}
}

\end{document}